\documentstyle[epsf]{mn}

\def\Rg{{R_{\rm g}}}
\def\Rin{{R_{\rm in}}}
\def\Rout{{R_{\rm out}}}
\def\Firr{{F_{\rm irr}}}
\def\mdcrit{{\dot m}_{\rm crit}}
\def\dm{{\dot m}}
\def\zycki{$\dot{\rm Z}$ycki}

\def\plotone#1{\centering \leavevmode
\epsfxsize=8.0truecm \epsfbox{#1}}
\def\plottwo#1#2{{\centering \leavevmode
\epsfxsize=8.0truecm \epsfbox{#1}} \centering \leavevmode
\epsfxsize=8.0truecm \epsfbox{#2}}

\begin{document}
 
\title[Relativistic Distortions in the X--ray Spectrum of Cyg X--1]
{Relativistic Distortions in the X--ray Spectrum of Cyg X--1}
\author[C. Done \& P.T. \zycki ]
{C. Done$^1$, P.T. \zycki$^{1,2}$ \\
$^1$ Department of Physics, University of Durham, South Road, Durham, DH1 3LE\\
$^2$ Nicolaus Copernicus Astronomical Center, Bartycka 18,
            00-716 Warsaw}

\maketitle
 
\begin{abstract}

We present the first significant detection of relativistic smearing of the
X--ray reflection spectrum from the putative accretion disk in the low/hard
state of Cyg X--1.  The ionization state, and amount of relativistic smearing
are simultaneously constrained by the X--ray spectra, and we conclude that the
disk is not strongly ionized, does not generally extend down to the last stable
orbit at 3 Schwarzschild radii and covers rather less than half the sky as seen
from the X--ray source. These results are consistent with a geometry where the
optically thick disk truncates at a few tens of Schwarzschild radii, with the
inner region occupied by the X--ray hot, optically thin(ish) plasma. Such a
geometry is also inferred from previous studies of the reflected spectrum in Galactic
Black Hole transient sources, and from detailed considerations of the overall
continuum spectral shape, suggesting that this is a robust feature of low/hard
state accretion onto Galactic Black Holes.

\end{abstract}

\begin{keywords}
accretion, accretion discs -- binaries: close -- black hole physics -- stars:
individual (Cygnus X-1) -- X--ray: general -- X--ray: stars
\end{keywords}

\section{INTRODUCTION}

Some of the strongest evidence for the existence of black holes has come from
recent ASCA (0.6--10 keV) X--ray observations of the shape of the iron K$\alpha$
line in Active Galactic Nuclei (AGN). AGN typically produce copious hard X--ray
emission, and the iron line is formed from fluorescence as these X--rays
illuminate the infalling material. The combination of Doppler
effects from the high orbital velocities and strong gravity in the vicinity of a
black hole gives the line a characteristically skewed, broad profile (Fabian
et al., 1989). This has been unambiguously identified from ASCA 0.6--10 keV
spectra of the AGN MCG--6--30--15, where the breadth of the line profile implies that
the accretion disk extends down to {\it at least} $6\Rg$, the last stable orbit
in a Schwarzschild metric (Tanaka et al 1995; Iwasawa et al 1996).

As well as producing the line, some fraction of the illuminating hard X--rays
are reflected from the accretion flow, producing a characteristic continuum
spectrum.  The amplitude of the line and reflected continuum depend on the
amount of material being illuminated by the hard X--rays, its inclination,
elemental abundances and ionization state (e.g.\ Lightman \& White 1988; George
\& Fabian 1991; Matt, Perola \& Piro 1991; Ross \& Fabian 1993; \zycki\ \& Czerny
1994). The mean amount of reflection and line seen in Seyfert 1 AGN is
consistent with a power law X--ray spectrum illuminating an optically thick,
(nearly) neutral disk, which subtends a solid angle of $\sim 2\pi$ (Pounds et al
1990; Nandra \& Pounds 1994), although it is now recognised that individual
objects show significant dispersion about this mean (Zdziarski, Lubi\'nski \& Smith
1999).

This contrasts with the situation in the Galactic Black Hole Candidates
(GBHC). These are also thought to be powered by accretion via a disk onto a
black hole, and, in their low/hard state, show spectra which are rather similar
to those from AGN. However, the amount of reflection and iron line is much less
than would be expected from an accretion disk which subtends a solid angle of
$2\pi$ (e.g. Gierli\'nski et al 1997a), and the detected line is narrow, with no
obvious broad component (Marshall et al 1993, Ebisawa et al 1996, hereafter
E96). There are (at least) two possible explanations for this: firstly that it
is an artifact of the difference in ionization state of the disk between GBHC
and AGN, or secondly that there is a real difference in geometry between the
supermassive accreting black holes and the stellar mass ones.

Ionization differences seems at first sight to be an extremely attractive
option. For accretion at the same fraction of Eddington, the disk temperature
should scale as $M^{-1/4}$. The GBHC inner disk is then expected to
be a factor of $\sim 30$ hotter than in AGN, and the higher expected
ionization state from the thermal ion populations
gives a reflected continuum and associated iron line that can
be very different to that from a neutral disk (Lightman \& White 1988; Ross \&
Fabian 1993; Matt Fabian \& Ross 1993ab; \zycki\ et al 1994; Matt Fabian \& Ross
1996; Ross, Fabian \& Brandt 1996). As well as increasing the energy of the iron
edge (and line), ionization also reduces the low energy opacity of the disk, and
so enhances its reflectivity. However, the iron edge (whatever its energy) has a
fairly constant cross-section, so the net result is to increase the relative strength of
the iron edge feature in the reflected continuum.  There is also the possibility
of the line being suppressed through resonant absorption over a (rather small)
range in ionization states, so the net result can be a prominent edge without
obvious line emission accompanying it. Such models were proposed as the solution
to the weak line emission seen in GBHC compared to AGN (Ross \& Fabian 1993;
Matt et al., 1993a; Ross et al 1996).

However, detailed spectral fitting with photo--ionized reflected continuua shows
that the ionization state of the reflector is in general too low for the
resonant absorption process to be important. This argues against such models,
although relativistic shifts could affect the derived ionization states (Matt
et al 1993a, Ross et al., 1996). Another way to suppress both
line and reflected continuum is from a radial dependance of ionization state. If
the inner region of the disk were so highly ionized that they produce no atomic
features then their contribution to the reflected continuum is unobservable in
the 2--20 keV range.  Here we fit data from Cyg X--1 with a reflection model
which includes both ionization (as a function of radius) and relativistic
smearing in order to test whether the spectra are indeed affected by resonant
absorption and whether radial ionization structure is important in masking the
reflection signature.

We significantly detect relativistic smearing of the reprocessed features in Cyg
X--1, and find that the disk is not highly ionized. This apparently rules out
both featureless (extreme ionization) reflection and resonant absorption.  The
line is not suppressed relative to the reflected continuum, it is merely
broadened by the relativistic effects which made it difficult to detect in ASCA
data (E96), where the observed weak and narrow line emission probably comes from
the companion star.  The derived covering fraction of the relativistic reflector
is substantially less than expected from a flat disk. The most plausible
explanation for this is that the optically thin(ish) X--ray corona fills a
central `hole', so that less than half of the X--ray flux intercepts the
disk. The amount of relativistic smearing generally is inconsistent with the
disk extending down to the last stable orbit at $6\Rg$.  These results from the
shape of the reprocessed spectrum are remarkably similar to the geometry derived
from energetic arguments from the continuum spectral shape (Gierli\'nski et al.\
1997a; Dove et al 1997; Poutanen, Krolik \& Ryde 1997).  Equally, they are very
similar to the derived reflection geometry from spectra of transient GBHC during
their decline (\zycki, Done \& Smith 1997; \zycki, Done \& Smith 1998ab), pointing
to a single, fairly stable geometry for the low/hard state GBHC.

\section{THE SPECTRAL MODEL}

The reflection model is described in detail in \zycki\ et al (1998b). Basically it consists
of the angle dependant reflection code of Magdziarz \& Zdziarski (1995), with
ion populations calculated as in Done et al (1992), as implemented in the
`pexriv' model in XSPEC. This determines the ion populations by balancing
photoelectric absorption against radiative recombination.  The photo--electric
absorption edge energies for iron were corrected from the rather approximate
values given by Reilman \& Manson (1978) to those of Kaastra \& Mewe (1994). As
noted by E96, these give a significant difference in the derived ionization
state. Photo--ionization rates are strongly dependent on the parameter $\xi =
L/(nr^2)$ ergs s$^{-1}$ cm$^{-1}$
(where L is the ionizing luminosity, here taken to be between 5eV and
20 keV, $n$ is the density and $r$ the distance of the material from the
ionizing flux) and weakly on spectral shape (assumed here to be a power law),
while recombination depends on temperature (assumed fixed at $10^6$ K). The
self--consistent iron line is then calculated from the Monte--Carlo code of
\zycki\ \& Czerny (1994) and added to the continuum. We assume fixed 'solar' abundances
from Morrison \& McCammon (1983), except for iron, which is free to vary from
its tabulated value of Fe/H $=3.3\times 10^{-5}$. 

Ionization as a function of radius can easily be treated by adding
together many radial rings with ionization $\xi(r)\propto r^{\beta}$, where the
relative covering fraction for each ring can be calculated from the X--ray
irradiation $\Firr(r)\propto r^{-3}$.

The relativistic smearing is taken from the XSPEC `diskline' model, with slight
modifications to take light bending into account (Fabian et al 1989).  This is
parameterised by the inner and outer radius of the disk, $\Rin$ and $\Rout$, its
inclination and (again) the irradiation emissivity, $\Firr(r)\propto r^{-3}$. We
fix $\Rout=10^4\Rg$ (where $\Rg=GM/c^2$) and fit for $\Rin$, convolving the
relativistic transfer function with the total (line plus continuum) reprocessed
spectrum. For the multiple ionization state disk the relativistic transfer
function is calculated for each annulus, convolved with the reprocessed spectrum
from that ring and then co--added to get the total spectrum.

We assume a power law illuminating spectrum. This is an approximation to the
true shape of a Comptonised spectrum and it becomes a poor description at
energies close to the seed photon energy and close to the electron temperature.
Typically in low/hard state of Cyg X--1 the electron temperature is $\sim 100$ keV
(e.g. Gierli\'nski et al 1997a), so this will not give significant distortions to
the spectral range considered here, which is always $\le 20$ keV. 

The situation with regards the soft photons is less clearcut. The strongest
component of the soft X--ray continuum can be modelled as a blackbody or disk
blackbody with a temperature of $0.1-0.2$ keV (e.g. Ba{\l}uci\'nska \& Hasinger
1991; E96; di Salvo et al 1998). Again, this is far enough away from the energy
range considered in this paper ($\ge 2$ keV) for the distortions of the true
Compton scattered spectrum from a power law to be negligible.  However, there is
also a {\it second} soft component which extends to higher energies (E96, di
Salvo et al 1998). E96 modelled this in the ASCA data using a softer power law
below a break energy around 3 keV, but such a broken power law is rather
unphysical.  Perhaps a more realistic model is one where soft photons from the
accretion disk (at $\sim 0.1$ keV) are Comptonised by a rather cool population
of electrons with $kT\sim 1$ keV as well as by the hot electrons with $kT\sim
150$ keV.  The disk photons at $\sim 0.1$ keV are energetically more important,
so it seems probable that this component at $\sim 1$ keV does not contribute
significantly to the seed photons. 

The physical origin of the second soft excess is entirely unknown, and so is its
geometry with respect to the disk and hot plasma. It may or may not then produce
its own reflected component. We assume it does not, but note that even if it
does then its contribution to the crucial iron line and edge region will be
very small due to the low temperature of this component. 

In the following fitting we generally neglect data below 4 keV so as to avoid the
complexities of the soft excess, and fix the galactic column at $6\times
10^{21}$ cm$^{-2}$.

\section{CONSTANT IONIZATION REFLECTION}

\subsection{EXOSAT GSPC data}

The EXOSAT GSPC data from Cyg X--1 give some of the best spectra to date from
this object, with the broad 2--20 keV energy range of GINGA data but resolution
comparable to that of the ASCA GIS. We use the 5 GSPC spectra of Done et al.,
(1992). The residuals to the best fit simple power law for each GSPC spectrum are
shown in Figure 1, together with the residuals to a simple power law fit to the
GINGA--12 AGN spectrum of Pounds et al (1990).  Contrary to claims by Ross et al
(1996), Cyg X--1 is not dominated by the edge structure, but also shows a strong
excess at the iron line energy. Also, {\it both} line and edge appear equally
diminished in Cyg X--1 compared to AGN, although it is somewhat misleading to
compare these residuals directly since Cyg X--1 is strongly absorbed by the
interstellar medium and can have a substantial contribution at 2 keV from the
observed soft excess.

\begin{figure}
\plotone{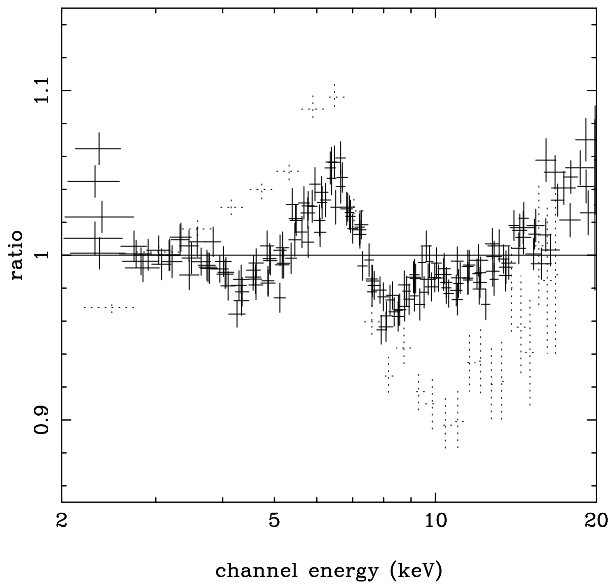}
\caption{
The ratio of each EXOSAT GSPC spectra of Cyg X--1 to a best fitting power law model. Both
excess counts at the line energy and a deficit at the edge are seen, as in AGN.
The dotted line shows the same ratio for the GINGA--12 AGN
spectrum (Pounds et al 1990) for comparison.
}
\end{figure}


We fit the 5 GSPC spectra simultaneously in XSPEC, although only the 
reflection parameters of inclination and iron abundance are constrained to be
equal across all the datasets. 

We first fit the 4--20 keV data with a power law.
This gives a poor fit ($\chi^2_\nu=1309/625$) due to the presence of
spectral features from iron (see Figure 1). Including a neutral reprocessed component
inclined at $30^\circ$ (Gies \& Bolton 1986)
gives a dramatic decrease in $\chi^2_\nu$ to $677/620$, with a
further significant improvement in the fit if this is allowed to be ionized
($\chi^2_\nu=611/615$). Allowing the iron abundance to be a free parameter makes
no significant difference, with $\chi^2_\nu=610/614$ for $A_{\rm Fe}=0.9\pm
0.2\times$ solar.

Allowing the reprocessed spectrum to be relativistically smeared gives a
significantly better fit, with $\chi^2_\nu=535/609$. We obtain constraints on
the inner radius of the disk in each spectrum, and these are somewhat larger
than the $6\Rg$ expected for a disk extending down to the last stable orbit, assuming an
illumination which is $\propto r^{-3}$. The derived iron abundance is $A_{\rm
Fe}= 1.8_{-0.4}^{+0.7}\times$ solar i.e. supersolar abundances are
preferred. This is consistent with the $\sim 2\times$ solar abundance of iron
(given as Fe/H$=6\pm 1\times 10^{-5}$)
inferred for the stellar wind of the companion star from the strong edge feature
seen in the absorbed `dip' spectra (Kitamoto et al., 1984). The relativistic
smearing allows more iron line to be present, since it is broad rather than
narrow, so it is much less observable. 

One potential problem may arise if there is a substantial narrow line component
from X--ray illumination of flared outer regions of the disk and/or the
companion star (Basko 1978) and stellar wind. This could distort the
relativistic line fit, by adding a sharp core to the line profile.  Such a line
is indeed detected in high resolution BBXRT and ASCA data (Marshall et al 1993;
E96).  We allow for this by including a narrow Gaussian line at 6.4 keV, whose
normalisation is a free parameter for each spectrum. This narrow component is
not significantly detected in the GSPC data, since it reduces $\chi^2$ by only
$\sim 1$ for 5 additional parameters.  The upper limits on the derived line
equivalent widths are typically less than $\sim 30$ eV, easily compatible with
the ASCA and BBXRT detections. Clearly there should also be a reflection
continuum accompanying this line unless a substantial portion of it arises from
an optically thin stellar wind. We model this by including an additional,
unsmeared, neutral reprocessed spectrum, assuming a fixed inclination of
$60^\circ$. Again, this does not improve the fit ($\chi^2_\nu=535/604$), 
but we include it for completeness. Details of this fit are given in Table 1,
where the iron abundance is frozen at its best fit value of $1.8\times$ solar so that the
error bars can be calculated for each spectrum separately. 


\begin{table*}
\begin{minipage}{180mm}
\caption{EXOSAT GSPC, GINGA and ASCA data from Cyg X--1. Error bars are $\Delta\chi^2=2.7$,
Galactic column is fixed at $6\times 10^{21}$ cm$^{-2}$, inclination at
$30^\circ$ and iron abundance at $1.8\times$ solar}
\label{}
\begin{tabular}{cccccccc}

dataset & PL $\Gamma$ & PL Norm & $\Omega/2\pi_{\rm rel}$ & $\xi$ & 
$R_{\rm in} (\Rg )$ & $\Omega/2\pi_{\rm non-rel}$ & $\chi^2_\nu$ \\
\hline
GSPC 08 & $1.75_{-0.01}^{+0.02}$ & 1.60 & 
$0.21^{+0.04}_{-0.05}$ & $17^{+40}_{-16}$ & $17^{+14}_{-8}$ & $0^{+0.07}$ & 89.2/118 \\
GSPC 09 & $1.56\pm 0.02$ & 1.24 &
$0.19^{+0.04}_{-0.07}$ & $24^{+90}_{-20}$ & $13^{+9}_{7}$ & $0^{+0.10}$ & 92.9/98 \\
GSPC 10 & $1.58\pm 0.02$ &  1.16 & $0.11^{+0.04}_{-0.06}$ & 
$30_{-26}^{+40}$ & $1000_{-952}$ & $0^{+0.12}$ & 69.3/104\\
GSPC 13 & $1.75^{+0.03}_{-0.02}$ & $2.35$ &
$0.24\pm 0.05$ & $31^{+60}_{-29}$ & $8_{-2}^{+12}$ & $0^{+0.11}$ & 174.8/176 \\
GSPC 14 & $1.71\pm 0.02$ & $1.75$ &
$0.22\pm 0.04$ & $46_{-42}^{+20}$ & $6.0^{+6}$ & $0^{+0.03}$ & 108.7/109 \\
\hline
GINGA 91-1 & $1.60^{+0.05}_{-0.07}$ & 1.61 & 
$0.23_{-0.11}^{+0.09}$ & $60_{-58}^{+640}$ & $10^{+\infty}_{-4}$ &
$0.07_{-0.07}^{+0.03}$\footnote{Parameter restricted to the range 0-0.1 since it
is not well constrained}& $10.2/21$ \\
GINGA 91-2 & $1.61^{+0.03}_{-0.08}$ & 1.85 & $0.29^{+0.06}_{-0.17}$ & 
$28^{+800}_{-28}$ & $28_{-22}^{+\infty}$ & $0^{+0.1}$$^a$ & 9.8/21\\
\hline
GIS 1 & $1.87\pm 0.02$ & $1.79$ & $0.05_{-0.01}^{+0.03}$ &
$2300_{-1600}^{+6000}$ & $16_{-5}^{+7}$ & $0^{+0.03}$ & 107.8/79\\
GIS 3 & $1.84_{-0.01}^{+0.02}$ & 1.80 & $0.06_{-0.02}^{+0.03}$ & 
$3\pm 2.4\times 10^4$ & $58_{-31}^{+110}$ & $0.04\pm 0.04$ & 92.0/79 \\
GIS 4 & $1.71\pm 0.02$ & 1.24 & $0.1\pm 0.03$ & 
$50_{-28}^{+100}$ & $90_{-40}^{+120}$ & $0^{+0.05}$ & 142.8/79 \\
SIS 5 & $1.68\pm 0.03$ & 1.22 & $0.09^{+0.78}_{-0.04}$ & 
$3^{+32}_{-2.2}\times 10^4$ & $23^{+38}_{-11}$ & $0.05^{+0.08}_{-0.04}$ & 51.8/70 \\
GIS 6 & $1.68\pm 0.02$ & 1.89 & $0.02_{-0.01}^{+0.09}$ & 
$6.70_{-0.01}^{+193}\times 10^3$ & $31_{-25}^{+210}$ & $0.13_{-0.05}^{+0.06}$ & 84.4/79 \\
GIS 7 & $1.44^{+0.03}_{-0.02}$ & 0.93 & $0.05_{-0.02}^{+0.08}$ &
$8.4_{-7.9}^{+42}\times 10^3$ & $14_{-6}^{+14}$ & $0^{+0.03}$ & 54.3/79 \\
GIS 8 & $1.66\pm 0.02$ & 1.44 & $0.06_{-0.04}^{+0.06}$ & 
$115_{-90}^{+1600}$ & $37_{-21}^{+40}$ & $0.05\pm 0.05$ & 108.2/79 \\
GIS 9 & $1.59\pm 0.02$ & 1.43 & $0.09^{+0.09}_{-0.08}$ &
$0^{+40}$ & $8_{-2}^{+36}$ & $0.13_{-0.06}^{+0.05}$ & 68.1/79 \\
GIS 10 & $1.60^{+0.03}_{-0.01}$ & 1.62 & $0.06_{-0.04}^{+0.11}$ &
$110_{-102}^{+2000}$ & $64_{-30}^{+100}$ & $0.08\pm 0.06$ & 95.8/79 \\

\hline

\end{tabular}
\end{minipage}
\end{table*}

Several points are immediately apparent from this series of models. Firstly, the
line strength is perfectly consistent with solar, or even supersolar
abundances. It is not anomalously weak compared to the reflected continuum as
proposed by Ross et al (1996). The reflector is significantly ionized, but not
to the strength that would be required for Auger ionization to operate ($\xi\sim
300-1000$ for a power law of $\Gamma=1.7$ and temperature of $10^6$ K). The
derived ionization state assuming an inclination of $30^\circ$ is typically that
of Fe VIII-XII, incompatible with the Auger range of Fe XVII -- XXIII.
These ionization states are obtained despite allowing for the effects of
relativistic smearing. There is no conspiracy, whereby the high Auger ionization
is masked by Doppler smearing and gravitational redshifts (Ross et al 1996).
Instead the line and reflection continuum are perfectly consistent with each
other, and with being produced in a disk which is weakly ionized.

\subsection{GINGA data}

The GINGA data give consistent results, though at much lower significance. There
were 3 epochs in which Cyg X--1 was observed, the first of which (August 1987)
was mispointed and the remaining two (in May 1990 and June 1991) being in an
unusual mode designed to increase the spectral range out to high energies. In
this mode the spectral resolution is reduced by about a factor 2 since the same
number of channels cover twice the bandpass. The resolution is then more like 30
per cent at iron, rather than 17 per cent in the usual mode.  This is a crucial
difference, since the broadest features expected from the inner disk are of
order 30 per cent if the disk extends down to the last stable orbit, so these
data are unlikely to be able to constrain the amount of relativistic
smearing. Ironically, the larger bandpass is not actually of use either, as data
beyond 30 keV are limited by uncertainties in background subtraction.

We use datasets 1 and 2 (the unabsorbed spectra) from the June 1991 observations
shown by Gierli\'nski et al (1997a). These are fit in the 4--30 keV range with the
same model as above.  The poor resolution of the data means that the
relativistic smearing is not significantly detected, so the relativistic and
non--relativistic reflected spectra are degenerate. From the EXOSAT results and
from theoretical expectations of the amount of reflection from the companion
star (Basko 1978) we limit the solid angle $\Omega/2\pi_{\rm non-rel}\le 0.1$.
With this constraint we obtain the results shown in Table 1. Again the data have
a line which is easily consistent with the amount of reflection seen. The
observed ionization parameter of the reflecting material is generally low (though
poorly constrained) even when relativistic distortions are included.  

\subsection{ASCA data}

The ASCA SIS and GIS data used here are those published by E96, except that we
neglect the October 1993 SIS spectra (spectrum 2 in E96) where there may be a
residual problem with the resolution (E96). Again we use only data above 4 keV
so as to avoid the complexities of the soft X--ray excess, and so that the 
fits can be compared with the results of E96. These data clearly show a narrow
line component as well as a complex edge structure which has been modelled by an
ionized reflection (E96). However, their reflected continuum model did not include
the self consistent iron line emission nor relativistic smearing.

The SIS spectrum has the best energy resolution (November 1993, denoted spectrum
5 in E96) so we first examine the line profile in these data.  We fit this with
the self--consistent iron line and continuum reflection code, together with a
power law continuum and galactic $N_H=6\times 10^{21}$, firstly assuming an
inclination of $30^\circ$. Without relativistic effects the fit is much worse
than that shown by E96, with $\chi^2_\nu=98/71$, showing that the narrow line
cannot be the corresponding fluorescence from the material giving rise to the
observed reflection continuum, irrespective of the iron abundance (best fit
value of unity) and ionization ($\xi\sim 5$).  Including relativistic effects
gives a much better fit, with $\chi^2_\nu=56/70$, comparable with that of E96.
Including a separate narrow line gives an even better fit, with line energy and
equivalent width of $6.46$ keV and $10$ eV ($\chi^2_\nu=48/68$). However,
physically we expect a self--consistent reflected continuum to accompany this
line. Replacing the line with the total reprocessed spectrum from neutral
material where relativistic effects are not important gives an equally good fit
with $\chi^2_\nu=51/69$. This fit (shown in Figure 2) is very different to the
one derived from the EXOSAT data. The ionization is so high ($\xi\sim 4\times
10^4$) that the `edge' feature seen in the data is actually modelled by the drop
following the relativistically smeared ($\Rin\sim 22\Rg$) ionized line, and the
real edge is shifted outside of the observed energy range.

\begin{figure}
\plotone{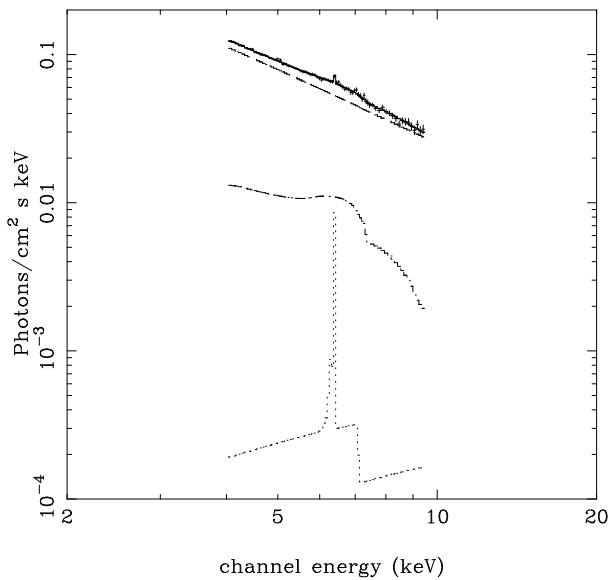}
\caption{
The best fit to the ASCA SIS 5 spectrum assuming an inclination of $30^\circ$,
with iron abundance free to vary between $1-2\times$ solar.
The relativistic reflector has $\xi=3.0_{-2.2}^{+37}\times 10^4$,
$\Rin=23_{-11}^{+40}$ and $\Omega/2\pi=0.09_{-0.04}^{+0.81}$, while the neutral
unsmeared reflector has $\Omega/2\pi=0.05_{-0.04}^{+0.10}$, for an illuminating
power law with $\alpha=0.68\pm 0.03$ ($\chi^2_\nu=51.8/70$). 
}
\end{figure}

The GIS spectra give similar results. We fit these using the most recent GIS response
matrices (as of March 1995, gis2v4-0.rmf) as opposed to the earlier response
used by E96, and results are detailed in Table 1. 
Again, the derived ionization paramter is generaly very high -- only the GIS 4 
and GIS 9 spectra give results that are anything like the low ionization EXOSAT
and GINGA solutions. 

\subsection{Inclination}

\begin{figure}\phantom{here}
\plottwo{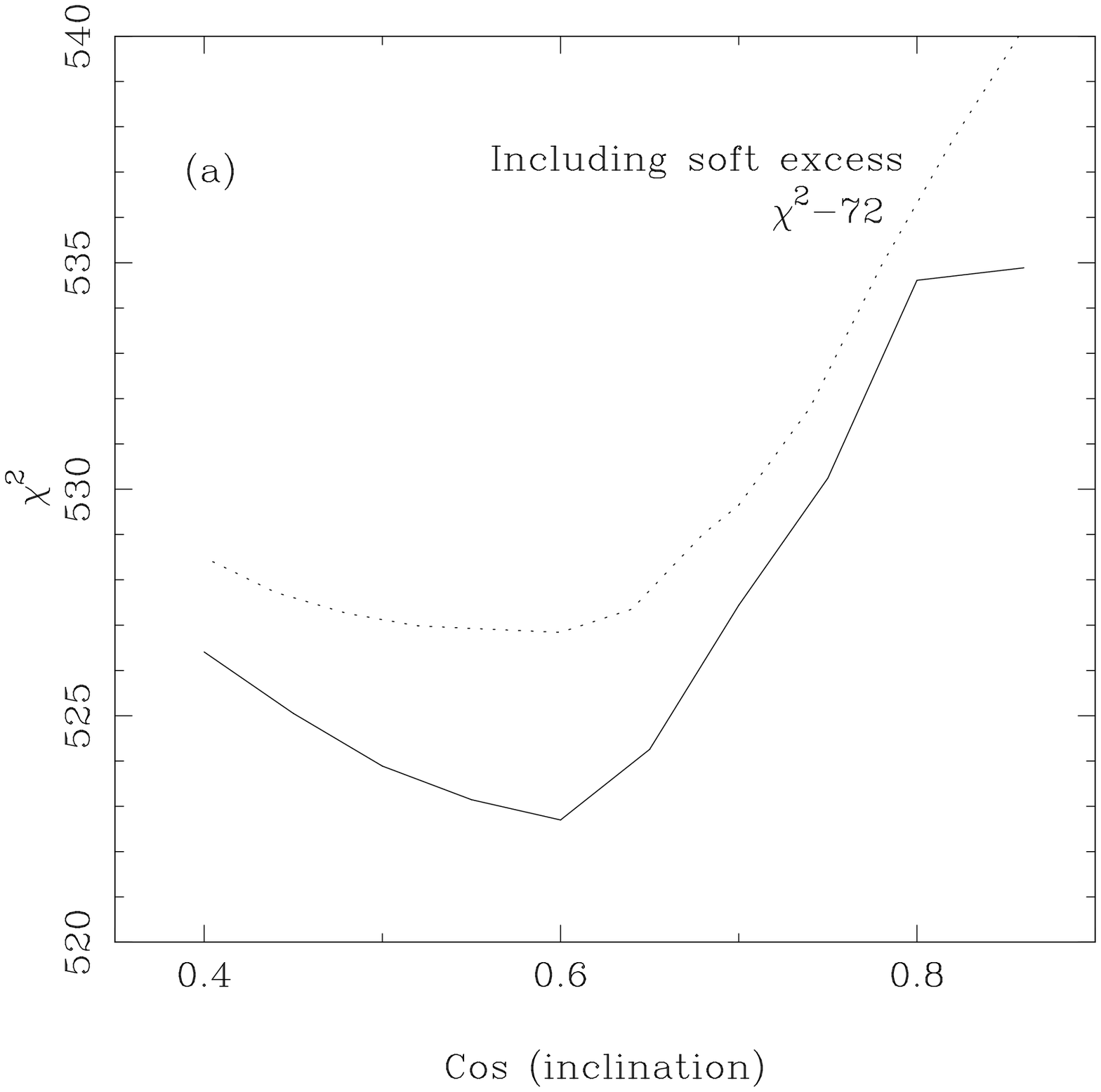}{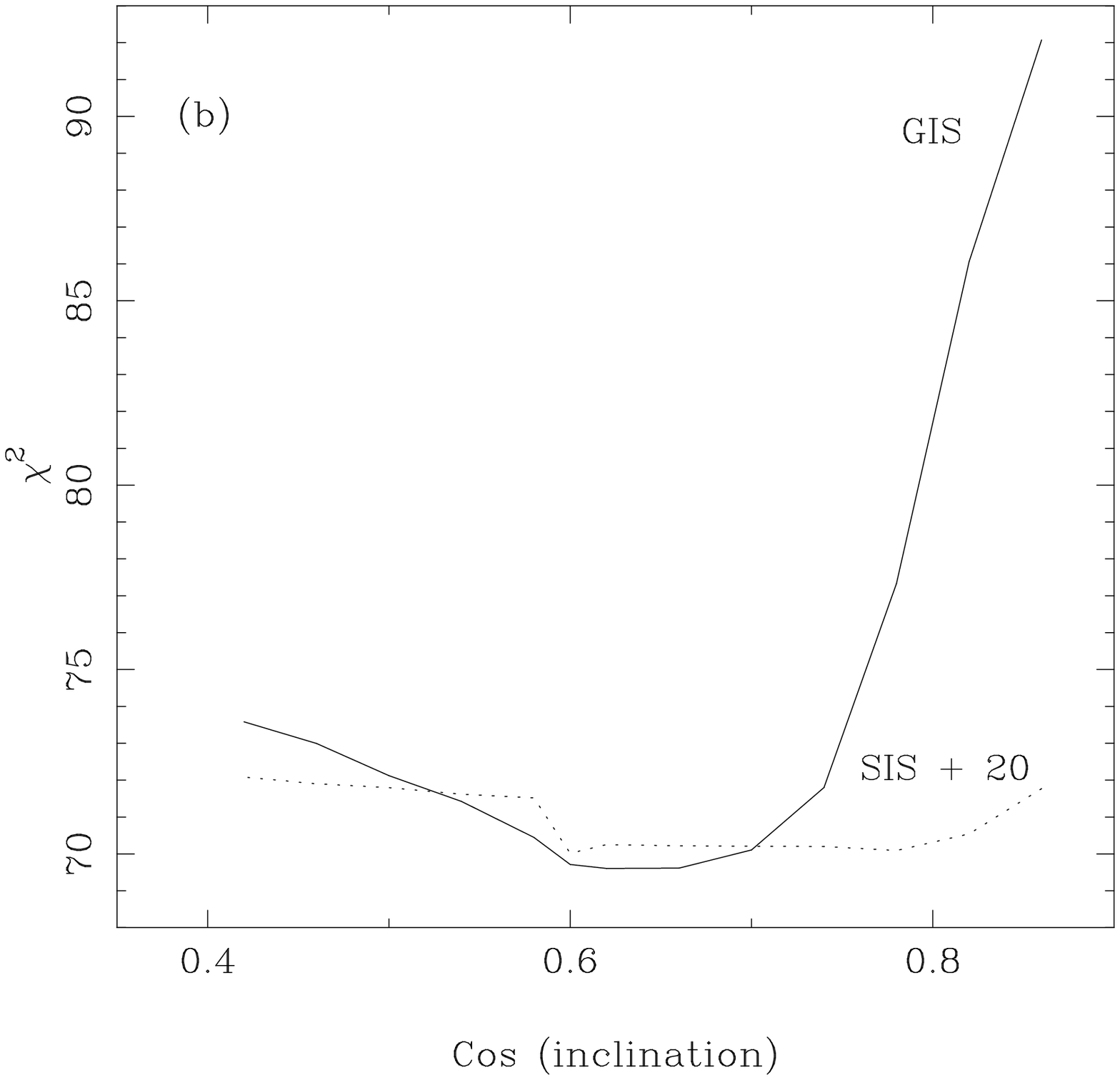}
\caption{
The change in $\chi^2$ (goodness of fit parameter) as a function of inclination
of the reflecting material.  Panel (a) shows results from a simultaneous fit 
of all the EXOSAT GSPC
spectra, with the solid line showing fits to the data from 4--20 keV while the
dotted line uses the full 2--20 keV bandpass including a 
(diskblackbody) model for the soft excess. The latter have 72 subtracted from
them in order to appear on the same scale. Panel (b) shows the ASCA GIS 3 (solid
line) and ASCA SIS 5 (dotted line) 4--10 keV spectral results, where the latter
have been boosted by 20 in order to appear on the same scale. Clearly
the data which have significant bandpass beyond 9 keV favour inclinations higher
than $30^\circ$.
}
\end{figure}

This general mismatch between the ionization of the reprocessor derived from
ASCA and EXOSAT/GINGA data is suggestive of a systematic problem in the spectral model
fitting, most likely due to the difference in bandpass. In EXOSAT and GINGA the high
energy continuum shape of the reprocessed spectrum helps constrain its
ionization, whereas in ASCA only the iron features can be used. The detailed
shape of the line and edge is a strong function of inclination as well as of
ionization. At high inclinations Doppler shifts prevail over gravitational and
transverse redshift, shifting the line and edge to higher energies as well as
giving substantial broadening. This is to zeroth order the same effect as 
ionization. However, the detailed shape of a low inclination, ionized 
reflection spectrum is rather different to a less ionized reflection spectrum at
higher inclination, so these two parameters can be disentangled
given good enough data. 

Figure 3a shows how $\chi^2$ varies for simultaneous spectral fitting of the
five EXOSAT datasets as the disk inclination is changed from $30-66^\circ$.
Clearly there is a significant preference in the data for an inclination higher
than $45^\circ$ (cosine smaller than 0.7). The 3$\sigma$ ($\Delta\chi^2=9$)
limit to the inclination from this modelling is $i\ge 40^\circ$. 
This is rather
higher than the most probable inclination inferred from optical studies of
$28^\circ-38^\circ$ (Gies \& Bolton 1986), but within the limit set by the most
recent work of $\le 55^\circ$ (Sowers et al 1998). At the best fit GSPC
inclination of $53^\circ$ the iron abundance is $2^{+0.6}_{-0.4}\times$ solar
($\chi^2_\nu=522/604$). We check that this is not an artifact of ignoring the
soft X--ray excess by including data down to 2 keV and approximating it with a
diskblackbody spectrum (XSPEC model {\tt diskbb}).  The dotted line in figure 3a
shows that the data still significantly favour the higher inclination solutions.

Despite their higher spectral resolution, the ASCA SIS are
completely unable to constrain the inclination (Figure 3b, dotted line).  Figure
4a shows this degeneracy of SIS solutions in more detail. At low inclinations
the data select uniquely the high ionization solution (shown in Figure 2), while
at high inclinations an equally good solution can be found at both high and low
ionization. Figures 5a and b show these high inclination solutions -- the low
ionization one has the 'pseudo--edge' in the data matched by the real edge while
at high ionization it is matched by the drop from the broadened ionized line.
Figures 5a and b also show the extension of these models out to 20 keV. It is
obvious that these solutions can be distinguished with higher energy data.

\begin{figure}
\plottwo{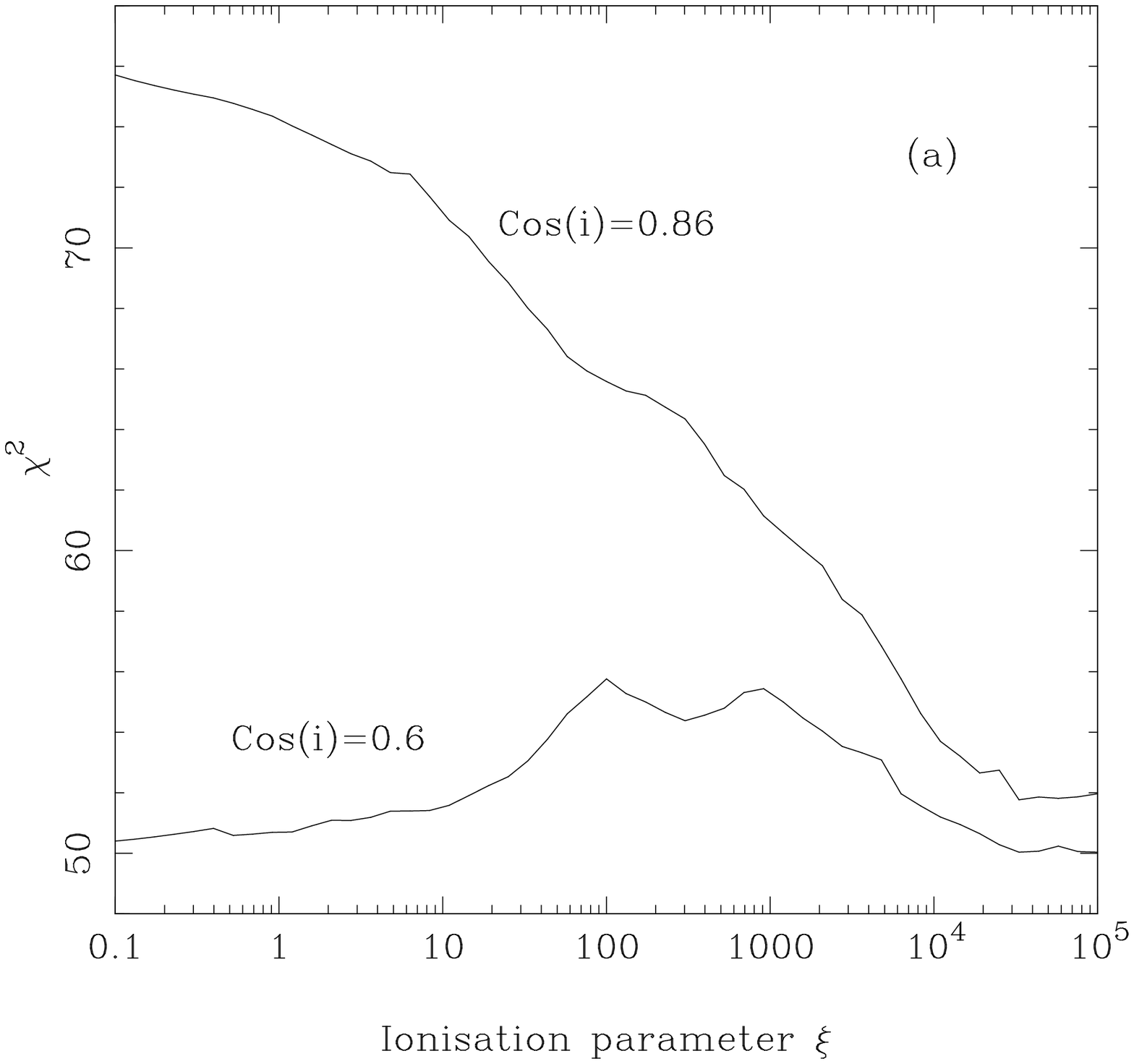}{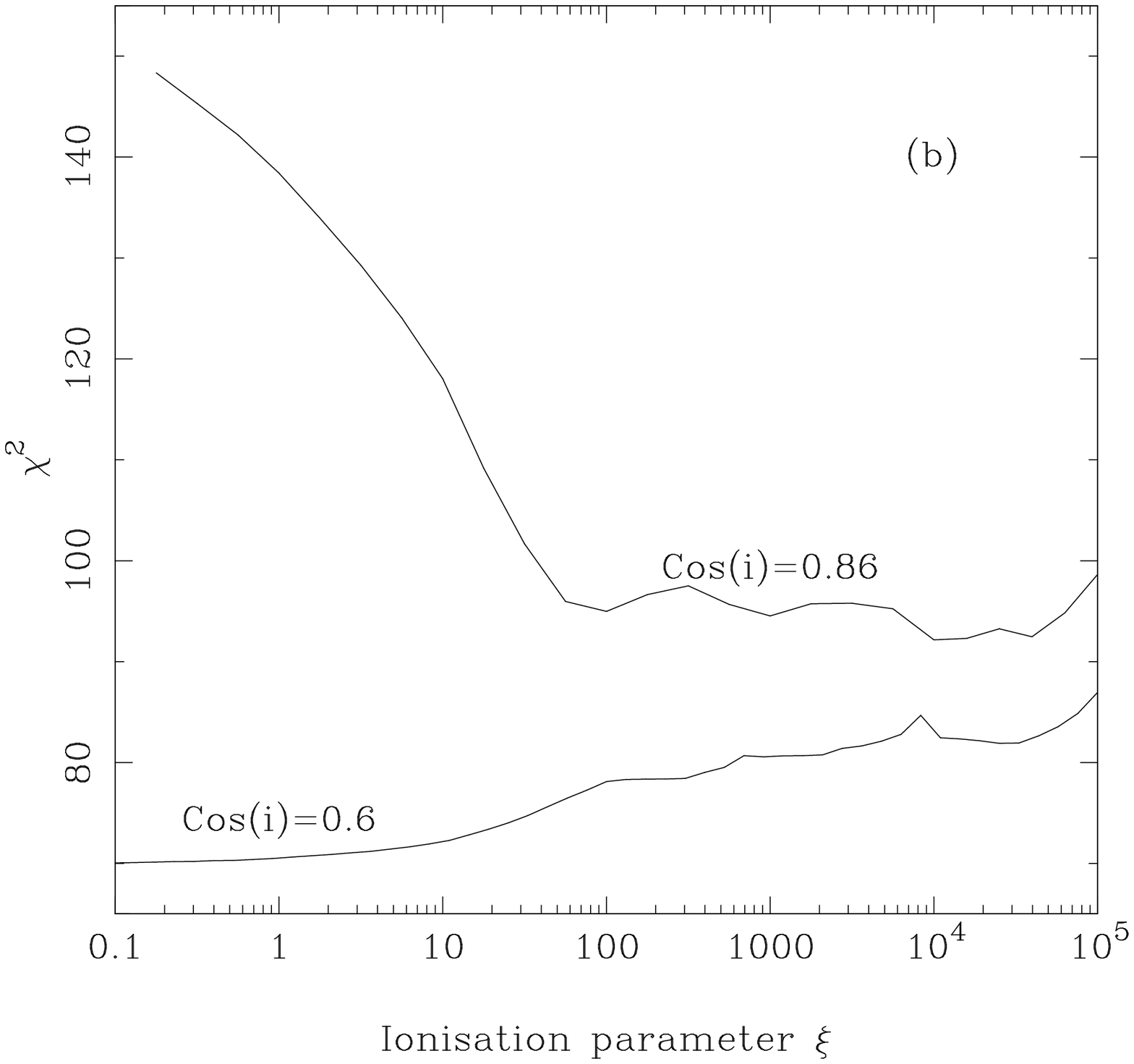}
\caption{
The change in the goodness of fit parameter $\chi^2$ as a
function of ionization state of the reflector for inclinations of $30^\circ$ and
$53^\circ$. Panel (a) shows the results for the SIS 5 spectrum. For an 
inclination of $30^\circ$ the data uniquely select the high ionization
solution while at $53^\circ$ there are equally good solutions at both low and
high ionization. All these three solutions are of comparable statistical
likelihood. Panel (b) shows the contrasting situation derived from the GIS 3
spectrum. Here the data uniquely select the high inclination, low ionization
solution. 
}
\end{figure}

\begin{figure}
\plottwo{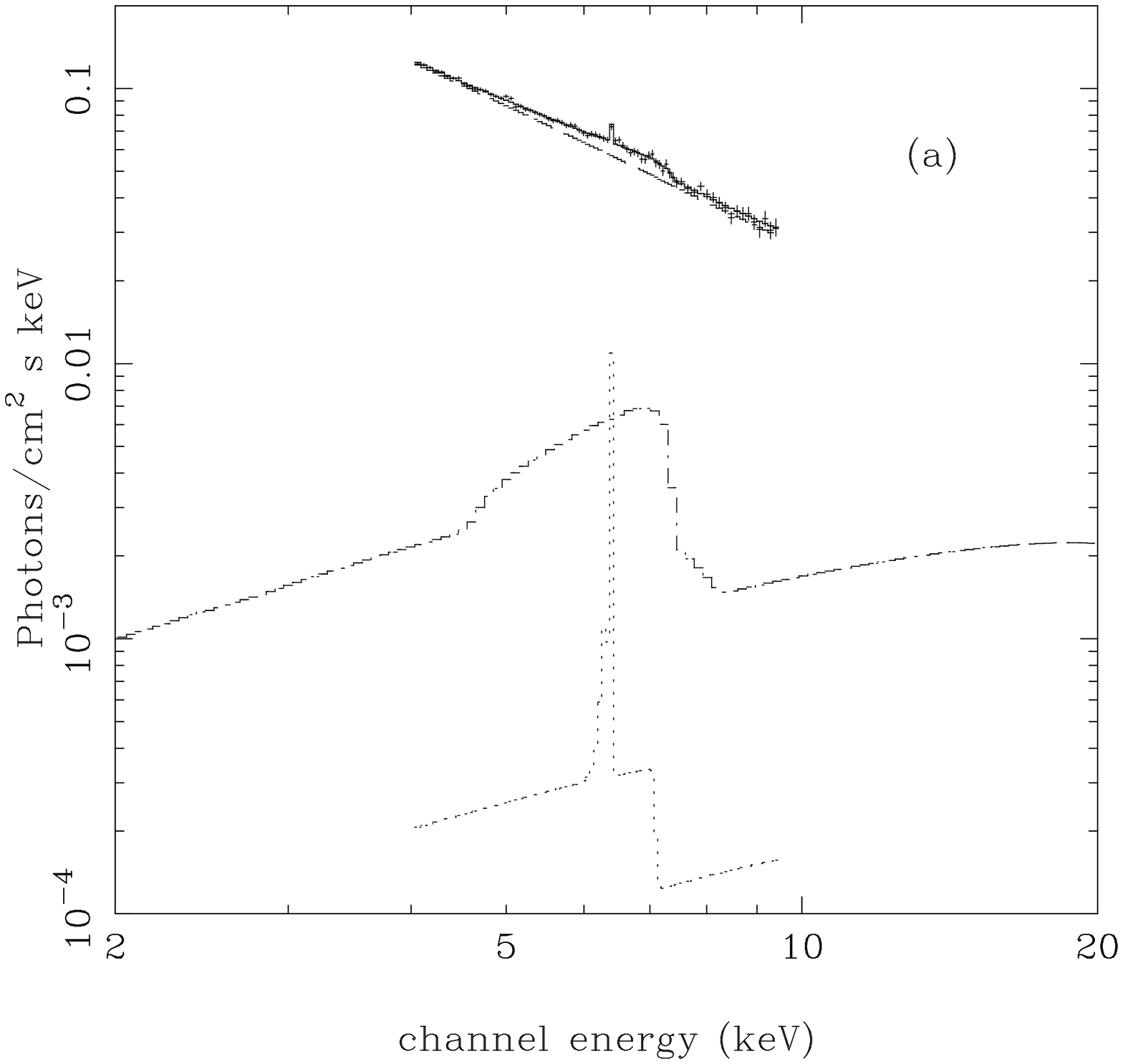}{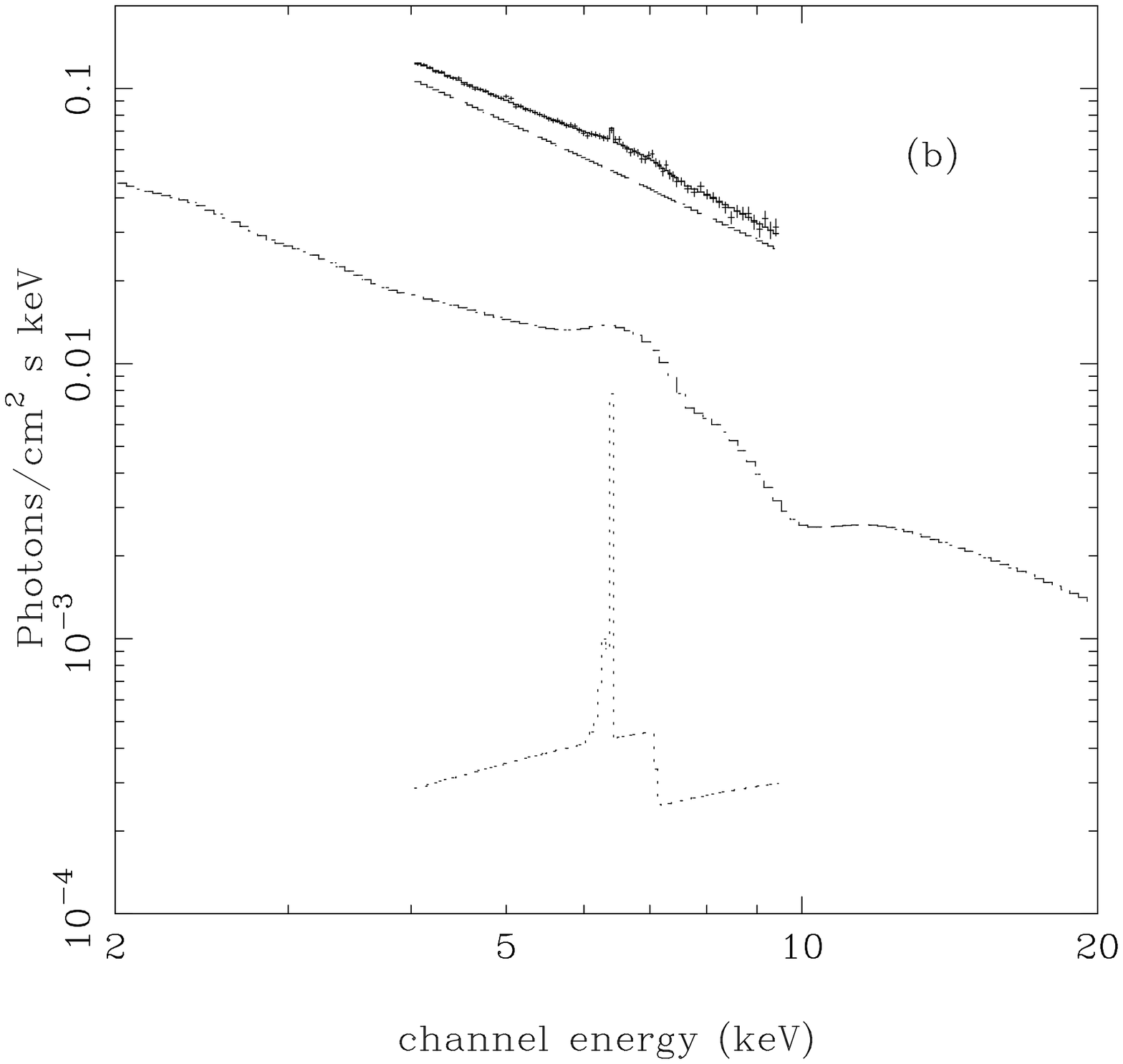}
\caption{
Two equally acceptable decompositions of the SIS 5 spectrum of Cyg X--1,
assuming an inclination of $53^\circ$ with iron abundance free to vary.
Panel (a) shows the low ionization
reflected continuum, with a relativistic disk of ionization
$\xi=0.04_{-0.04}^{+25}$, $\Rin=14\pm 4$ and $\Omega/2\pi=0.45_{-0.10}^{+0.80}$
($\chi^2_\nu=49.7/69$), for an illuminating power law with
$\alpha=0.73_{-0.04}^{+0.16}$ and neutral unsmeared reflection
$\Omega/2\pi=0.05^{+0.10}_{-0.03}$. The iron abundance is completely
unconstrained between $0.5-3\times$ solar.
Panel (b) shows the high ionization
solution, where the relativistic disk has $\xi=3^{+34}_{-2.6}\times 10^4$,
$\Rin=55^{+60}_{-30}$ and $\Omega/2\pi=0.18_{-0.11}^{+4.0}$, with $\Omega/2\pi=0.07\pm
0.06$ of unsmeared neutral reflection and a power law with $\alpha=0.70\pm 0.04$
($\chi^2_\nu=50.0/69$).
}
\end{figure}

The GIS data have just enough extra bandpass at high energies (above 9 keV) to
break this degeneracy. The dashed line in figure 4b shows the change in $\chi^2$
with ionization for $\cos i = 0.86$ and $0.6$ for the GIS spectrum 3 of E96. Not
only is the ionization uniquely determined for a given inclination, the data
also are able to distinguish between the high ionization/low inclination and low
ionization/high inclination solutions, clearly preferring the latter.  The
correspondence with the GSPC data is clear, so we fix the inclination of the
relativistic reflector at $53^\circ$ (cosine of 0.6) and re--fit all the data,
using the best fit GSPC iron abundance of twice solar. These results are shown
in Table 2. For the EXOSAT GSPC (and GINGA) data the fits are only subtly
different -- the derived ionization state is somewhat lower while the inner
radius is somewhat larger, both because of the stronger doppler effects in the
higher inclination models. However, for the ASCA data the high inclination
solutions are very different from those in Table 1. Firstly, they are
statistically better -- the total $\chi^2$ from all the GIS data for the high
inclination solution is $698.5/632$ as opposed to $753.4/632$ for the low
inclination one. Secondly, the derived ionization drops dramatically, and
now corresponds to that seen in the EXOSAT GSPC data. All the data now give 
similar derived parameters, irrespective of bandpass. Only the SIS 5 and GIS 6
spectra allow a high ionization solution. 

Unlike the variable gain EXOSAT GSPC data, the GIS data are all taken in the
same mode, so their spectral residuals can be co--added in order to illustrate
the features which make one model better than another. The data are fit
simultaneously, with only the reflection parameters of iron abundance
(constrained to be between 1--2$\times$ solar) and
inclination fixed between the datasets. The top panel in Figure 6 shows the
results from fitting a reflector which is ionized but not relativistically
smeared for an inclination of $30^\circ$. There are clear broad residuals
present around the iron line energy. A much better fit is obtained if the
$30^\circ$ ionized reflector is allowed to be relativistically smeared (and a non--smeared,
unionized reflector is also added), as shown in the middle panel. However, the
bluest features of the residuals are not able to be adequately fit in these
models, despite the rather high derived ionizations (see table 1), due to the
predominant redshift at $30^\circ$. 
Higher inclination solutions give a stronger doppler shift to the line,
so matching the data rather better (bottom panel).

\begin{figure}
\plotone{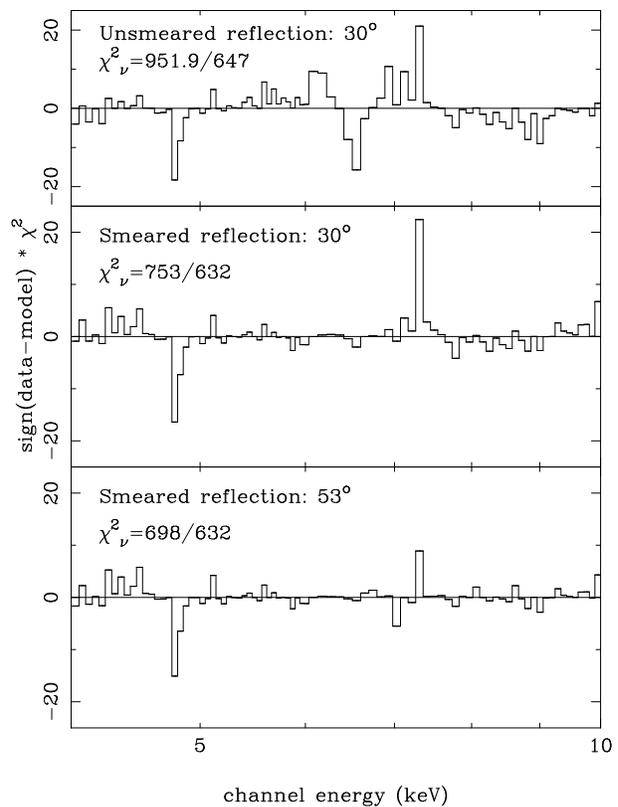}
\caption{
Co--added $\chi^2$ residuals from fitting the 8 GIS spectra. The top panel shows
results from fitting non--smeared reflection to the data. There are clearly
broad residuals left around the iron line energy. The mid panel shows that these
are much reduced by allowing the reflector to be a relativistic disk 
(including non--smeared, unionized reflection from the companion star). 
However, for the assumed inclination of $30^\circ$, the highest energy residuals
are not matched by the models. The lower panel shows that these can be fit at higher
inclinations, where doppler blueshifts can become stronger than the prevailing
redshifts. The feature at 4.7 keV in all the plots is a residual from the 
changing response of the GIS detectors at the Xe L edge.
}
\end{figure}

For the high inclination solutions, all the data show that
the covering fraction of the relativistic reflector is always much less than
unity, the derived inner disk radius is generally significantly larger than
$6\Rg$, and its ionization state is rather low. The solid angle subtended by the
narrow reprocessor is always consistent with being the $\Omega/2\pi
\le 0.1$ expected from the companion star. This should
vary with orbital phase (Basko 1978), although the data here give
error bars on the derived parameters which are too large to constrain this.
Orbital variability in the narrow reflected component would provide an explanation for
the observed 4\% peak to peak quasi--sinusoidal variability seen in the high
energy BATSE (20--100 keV) lightcurve of Cyg X-1 (Paciesas et al 1997).
The reflected component should be small (basically zero) at phase $0$ when the
companion star is in front of the black hole, and have its maximum of
$\Omega/2\pi \sim 0.1$ viewed at an inclination of $90-i$ at phase $0.5$,
giving a 5\% peak--to--peak variability in the 20--100 keV flux.
We do not require an additional narrow reflected component from the outer disk,
contrary to the conclusions of E96. Their fits to these data
gave some spectra which required a much larger narrow component
than could easily be explained by the companion star alone (e.g. GIS data from
November 1994, spectrum 10), even allowing for twice solar iron abundances. The
reason for the difference is that the fit is somewhat sensitive to changes in
the model, and we note that our description gives better $\chi^2$ (by up to 20)
than the fits of E96.

\begin{table*}
\begin{minipage}{180mm}
\caption{EXOSAT GSPC, GINGA and ASCA data from Cyg X--1. Error bars are $\Delta\chi^2=2.7$,
Galactic column is fixed at $6\times 10^{21}$ cm$^{-2}$, inclination is 
$53^\circ$ and the iron abundance is $2\times$ solar}
\label{}
\begin{tabular}{cccccccc}

dataset & PL $\Gamma$ & PL Norm & $\Omega/2\pi_{\rm rel}$ & $\xi$ &
$R_{\rm in}$ & $\Omega/2\pi_{\rm non-rel}$ & $\chi^2_\nu$ \\
\hline
GSPC 08 & $1.77\pm 0.02$ & 1.65 & $0.34^{+0.05}_{-0.14}$ 
& $0.2_{-0.2}^{+17}$ & $30_{-11}^{+18}$ & $0^{+0.14}$ & $86.1/118$ \\
GSPC 09 & $1.58^{+0.02}_{-0.01}$ & 1.28 & $0.26^{+0.09}_{-0.12}$ 
& $0^{+0.3}$ & $28^{+22}_{-15}$ & $0.05_{-0.05}^{+0.11}$ & $96.3/98$ \\
GSPC 10 &  $1.58^{+0.02}_{-0.01}$ &  1.16 & $0.13_{-0.07}^{+0.05}$ 
& $21_{-20}^{+70}$ & $1000_{-970}$ & $0^{+0.13}$ & $69.7/104$ \\
GSPC 13 & $1.78^{+0.02}_{-0.03}$ & $2.46$ & $0.34^{+0.12}_{-0.13}$ 
& $0^{+3}$ & $15_{-9}^{+15}$ & $0.09_{-0.09}^{+0.10}$ & $172.6/176$ \\
GSPC 14 & $1.74^{+0.01}_{-0.02}$ & $1.85$ & $0.39\pm 0.06$ & $0^{+0.4}$ 
& $14_{-4}^{+7}$ & $0^{+0.05}$ & $98.0/109$ \\
\hline
GINGA 91-1 & $1.63\pm 0.03$ & 1.60 & $0.33_{-0.09}^{+0.15}$ & 
$4_{-4}^{+30}$ & $40^{+\infty}_{-30}$ & 
$0.1_{-0.1}$\footnote{Parameter restricted to the range 0-0.1 since it is not well
constrained}& $12.1/21$ \\
GINGA 91-2 & $1.63\pm 0.03$ & 1.92 & $0.44^{+0.10}_{-0.21}$ & 
$3^{+45}_{-3}$ & $30^{+\infty}_{-22}$ & $0^{+0.1a}$ & 8.8/21\\
\hline
GIS 1 & $1.91\pm 0.02$ & 1.89 & $0.36_{-0.13}^{+0.12}$  
& $2.8_{-2.7}^{+40}$ & $20_{-5}^{+6}$ & $0.01_{-0.01}^{+0.04}$ & $93.5/79$ \\
GIS 3 & $1.91\pm 0.02$ & 2.08 & $0.53^{+0.09}_{-0.17}$ 
& $0.02_{-0.02}^{+17}$ & $9.9_{-1.9}^{+6}$ & $0.08_{-0.04}^{+0.03}$ & $69.8/79$\\
GIS 4 & $1.74\pm 0.02$ & 1.28 & $0.31\pm 0.08$ 
& $0^{+1}$ & $6.0^{+2}$ & $0.09\pm 0.03$ & $134.4/79$\\
SIS 5 & $1.74\pm 0.04$ & 1.46 & $0.54_{-0.13}^{+0.17}$ 
& $0^{+20}$ & $13_{-3}^{+4}$ & $0.05\pm 0.03$ & $49.9/70$ \\
GIS 6 & $1.69^{+0.03}_{-0.02}$ & 1.94 & $0.10^{+0.17}_{-0.09}$
& $30_{-30}^{+1e5}$ & $18^{+30}_{-12}$ & $0.12\pm 0.03$ & $81.4/79$ \\
GIS 7 & $1.50\pm 0.03$ & 1.01 & $0.48^{+0.07}_{-0.11}$
& $0^{+10}$ & $14^{+5}_{-4}$ & $0^{+0.02}$ & $45.7/79$ \\
GIS 8 & $1.66_{-0.01}^{+0.02}$ & 1.45 & $0.09_{-0.07}^{+0.05}$ 
& $50_{-45}^{+2000}$ & $60^{+170}_{-44}$ & $0.05_{-0.04}^{+0.05}$ & $108.4/79$ \\
GIS 9 & $1.58\pm 0.02$ & 1.42 & $0.04^{+0.13}_{-0.04}$ 
& $0^{+\infty}$ & $51_{-45}^{+\infty}$ & $0.12_{-0.12}^{+0.04}$ & $71.7/79$\\
GIS 10 & $1.61\pm 0.02$ & 1.63 & $0.04^{+0.09}_{-0.02}$ 
& $700_{-680}^{+1700}$ & $100^{+220}_{-66}$ & $0.09_{-0.05}^{+0.04}$ & $93.6/79$
\\

\hline

\end{tabular}
\end{minipage}
\end{table*}

\begin{table*}
\begin{minipage}{180mm}
\caption{GSPC 08, SIS 5 and GIS 3 spectra fit to the radial ionization model
with $\xi(r)\propto r^{-1.5}$.
Error bars are $\Delta\chi^2=2.7$, Galactic column is fixed at $6\times 10^{21}$
cm$^{-2}$. Inclination is fixed at $30^\circ$ and $53^\circ$ for each spectrum,
while the iron abundance is allowed to vary between 1--2$\times$ solar (so there
is an extra degree of freedom compared to table 2).}
\label{}
\begin{tabular}{cccccccccc}
 
Spectrum & PL $\Gamma$ & $A_{\rm Fe}$ & Cos(i) &
$\Omega/2\pi_{\rm rel}$ & $\xi(R_{\rm in})$ & $R_{\rm in}$ & 
$\Omega/2\pi_{\rm non-rel}$ & $\chi^2_\nu$ \\
\hline
GSPC 08 & $1.77^{+0.01}_{-0.03}$ & 2.0 & 0.60 & $0.34^{+0.04}_{-0.14}$ & 
$2_{-2}^{+50}$ & $27_{-8}^{+17}$ & $0.01^{+0.12}_{-0.01}$ & 86.2/117\\
GSPC 08 & $1.77\pm 0.04$ & 1.2 & 0.86 & $0.22_{-0.03}^{+0.04}$ & 
$65^{+160}_{-56}$ & $20_{-7}^{+13}$ & $0^{+0.09}$ & 88.5/117 \\
GIS 3 & $1.91^{+0.09}_{-0.02}$ & 2 & 0.60 & $0.53^{+0.40}_{-0.11}$ & 
$0.01^{+11}_{-0.01}$ & $9.5^{+3}_{-2.5}$ & $0.09^{+0.07}_{-0.04}$ & 69.7/78 \\
GIS 3 & $1.85^{+0.01}_{-0.02}$ & 2 & 0.86 & $0.06\pm 0.01$ & 
$3.0_{+3.5}^{-2.1}\times 10^4$ & $34^{+27}_{-12}$ & $0^{+0.06}$ & 89.7/78 \\
SIS 5 & $1.75^{+0.13}_{-0.04}$ & 2 & 0.6 & $0.56^{+0.59}_{-0.12}$ &
$0.1_{-0.1}^{+30}$ & $13_{-3.5}^{+6}$ & $0.06_{-0.03}^{+0.08}$ & 49.4/69\\ 
SIS 5 & $1.69^{+0.05}_{-0.03}$ & 2 & 0.6 & $1.08^{+\infty}_{-0.99}$ &
$1.40_{-1.37}^{+2.1}\times 10^6$ & $17_{-10}^{34}$ & $0.11^{+\infty}_{-0.06}$ & 50.7/69 \\ 
SIS 5 & $1.67^{+0.05}_{-0.03}$ & 2 & 0.86 & $0.11^{+\infty}_{-0.06}$ & 
$1.8^{+38}_{-1.5}\times 10^5$ & $14_{-6}^{+16}$ & $0.05^{+\infty}_{-0.04}$ & 52.0/69\\ 
\hline

\end{tabular}
\end{minipage}
\end{table*}

\section{IONIZATION AS A FUNCTION OF RADIUS}

The data clearly prefer a solution in which the covering fraction of the
reflector is extremely low, typically only 20--30 per cent of what would have
been expected from a disk which covered half the sky as seen from the X--ray
source, and the derived inner radius is generally inconsistent with the last
stable orbit. Both these results can easily be explained in a scenario where the
optically thick disk terminates before the last stable orbit, with the inner
radii instead being filled with X--ray hot optically thin(ish) plasma. 
However, a series of papers (e.g. Ross et al 1996) have explored the
possibility that high ionization of the reflector supresses the derived amount
of reflection. We have shown above that this does not work for Cyg X--1 when the
reflector is modelled by material with constant ionization parameter. 
However, perhaps there is a radial range of ionization states. If
there is an extremely ionized inner disk then it produces
no spectral features, so neither its covering fraction nor its relativistic
smearing can be seen. 

We test this idea by dividing the disk into 10 radial zones,
again with the illumination $\propto r^{-3}$, and the ionization $\xi(r) \propto
r^{-1.5}$ as appropriate for a gas pressure dominated disk. Due to computational
time we only demonstrate this model on a single spectrum from each instrument:
EXOSAT GSPC 08, ASCA GIS 3 and  ASCA SIS 5. The iron abundance is a free
parameter (within the range $1-2\times$ solar), and we include the possibility
of a neutral reflected component from the star/outer disk. The fits are repeated
for fixed inclination of $\cos i = 0.86$ and $0.6$. 

Details are given in Table 3, and are not significantly different from those
derived from the single ionization model.  The inferred covering fraction for
the disk is still low, and the disk inner radius is inconsistent with the last
stable orbit.  We explicitally searched for a high ionization solution for the
$\cos i = 0.6$ fits, but these always gave a significantly worse $\chi^2$ except
for the SIS 5 data.

The reason for the failure of the multi--zone ionization model is that the mean
ionization state observed is rather low.  Between this and the extreme
ionization states required to make the reflector invisible are the
intermediate/high ionization states. These give a strong and distinctive
spectral signature, especially those where iron has only one or two electrons
left (H and He--like, FeXXVI and XXV) which produce intense line emission at 7.0
and 6.7 keV, and (more importantly) a deep edge at 9.2 and 8.6 keV (Ross \&
Fabian 1993).  Even allowing for strong distortion by relativistic smearing
(Matt et al., 1993a; 1996; Ross et al., 1996) these features are not seen in the
data. It is not then possible to go smoothly from the observed rather low
ionization states to the extreme ionizations required to hide reflection. Only
if there is an abrupt ionization transition can this model be sustained.

Again, we note that good data at energies $\ge 9$ keV are required in order to
constrain the amount of highly ionized reflecting material. Radial ionization
models with high inner disk ionization can be fit to the SIS data for any
inclination, but are again ruled out by the GIS 3 spectrum (see table 3).

\section{DISCUSSION}

From the above fits it is clear that the relativistic smearing of the reflected
spectrum is significantly detected in Cyg X--1, confirming the presence of an
optically thick material at small radii -- the putative accretion disk. The
covering fraction of the disk is substantially less than unity (see Tables 1-3),
irrespective of inclination, ionization and radial gradients in ionization. 

The ionization of the reflector and its inclination to the line of sight can
be constrained with moderate spectral resolution data {\it and} bandpass
extending beyond 9 keV. The data show strong preference for a low ionization
reflector inclined at rather more than $30^\circ$ to the line of sight: the ASCA
GIS 3 spectrum gives a range of $41^\circ - 61^\circ$, while the EXOSAT GSPC
data imply angles greater than $\sim 47^\circ$ (both at the 90\% confidence
limit). These inclinations are determined by the detailed shape of the iron line
in the data: for low inclinations the theoretical line profile is strongly
affected by gravitational redshift so that higher ionization solutions are
needed to fit the data. But in higher ionization solutions there is a large
energy separation between the line and edge (however smeared they become) which
is {\it not} seen in the data.  The spectral features observed require that the
line and edge blend together, i.e. that the ionization is moderate to low. The
reflector then is required to be viewed at a moderate inclination angle so that
the doppler blueshift can counteract the gravitational redshift.  However, these
inclination are rather larger than the often quoted probable range of
$28^\circ-38^\circ$ (Gies
\& Bolton 1986) derived from optical data, although within the
firm upper limit to the inclination from these studies of $i\le 55^\circ$
(Sowers et al 1998). The lack of a well determined optical orbital solution is
due to the complexity of the accretion flow from the supergiant companion in
this system. The companion probably does not completely fill its Roche lobe, but
instead has a strong stellar wind which is gravitationally focussed by the Roche
potential (see e.g. numerical simulations by Blondin, Stevens \& Kallman 1991)

While the iron line is strongly distorted by the relativistic effects, the
extremely broad components from material at $6\Rg$ are not present at the level
expected from a disk extending down to the last stable orbit with line
emissivity $\propto r^{-3}$. Such a model is only possible if the inner disk has
an {\it abrupt} increase in ionization state to the level at which it becomes
completely ionized so produces no atomic spectral features. We stress that a
continuous (power law) ionization gradient as expected from the intense X--ray
illumination is insufficient to do this, since between the observed rather low
ionization state of the reflector and the extreme ionizations required to make
the disk completely reflective there are the high ionization states, where iron
has only 1 or 2 electrons left. These ions produce copious K$\alpha$ line
emission at 6.7 and 7.0 keV, and strong edges at 8.8 and 9.2 keV, which are not
seen in the data. Similarly, these high ionization states are precisely the ones
expected from collisional ionization from the accretion disk temperature. The
observed soft excess is dominated by a component at $0.1-0.2 $ keV which cannot
completely ionize iron. Even the secondary soft emission at $kT\sim 1$ keV is
far removed from the inner K shell energy of 8.8 and 9.2 keV for He and H like
iron.  This then rules out the idea of a simple correspondence between AGN and
GBHC where both have disks which extend down to $6\Rg$ but in the GBHC the disk
is ionized simply because of the higher accretion disk temperature.

The optically thick disk must do {\it something} at the derived $R_{\rm in} >
6\Rg$ (assuming that the line emissivity is $\propto r^{-3}$; \zycki\ et al
1998b). Either the inner disk disappears
completely or it jumps {\it discontinuously} to a completely ionized state at
this radius. Similar results are seen for the low/hard state GBHC transient
sources
(\zycki\ et al 1997; 1998ab), showing that this is a robust feature of low/hard state
GBHC spectra. 

Figure 7 shows the plot of spectral index versus the derived covering fraction,
ionization and inner radius of the reflector for the reflector models inclined
at $53^\circ$ (not including the poorly constrained GINGA data).  The covering
fraction is not consistent with a constant ($\chi^2_\nu=60/14$), showing that
the geometry must be changing. Apart from the hardest spectral index data (GIS
7, which incidentally is the spectrum taken furthest off axis, and so has a
large vignetting correction, E96) the data are consistent with a trend in which
a hard spectral index is associated with smaller reflected fraction, similar to
results from some of the X--ray transients (\zycki\ et al 1997; 1998ab). This
can be qualitatively explained by a geometry in which there is a variable radius
inner hole in the cool disk which is filled with X--ray hot plasma, although the
error bars on the radius derived from relativistic smearing are too large to be
able to constrain these ideas.  A disk which extends further down into the
gravitational potential could subtend a larger solid angle to the source. The
soft photon flux from the disk would also increase, both because the disk
luminosity is dependent on its inner radius, and also from the larger fraction
of hard X--rays which can be reprocesssed. This would give stronger Compton
cooling of the hot electrons and so a steeper X--ray spectrum. 

The typical derived inner radius of the disk is $R_{\rm in}=20\Rg$.  The
luminosity released by viscous dissipation from $20\Rg$ to $6\Rg$ is rather less
than $0.75$ that released from $\infty$ to $20\Rg$ (see e.g. equation 5.19
in Frank, King \& Raine 1992). Simple energetics then give the expectation that
the X--ray hot plasma should carry $\sim 0.75\times$ the luminosity of the
soft photons from the disk. However, the opposite is observed:
the hard component (the X--ray hot plasma) is
over twice as luminous as the soft component from the disk e.g. di Salvo et al.,
(1998). 

\begin{figure}
\plotone{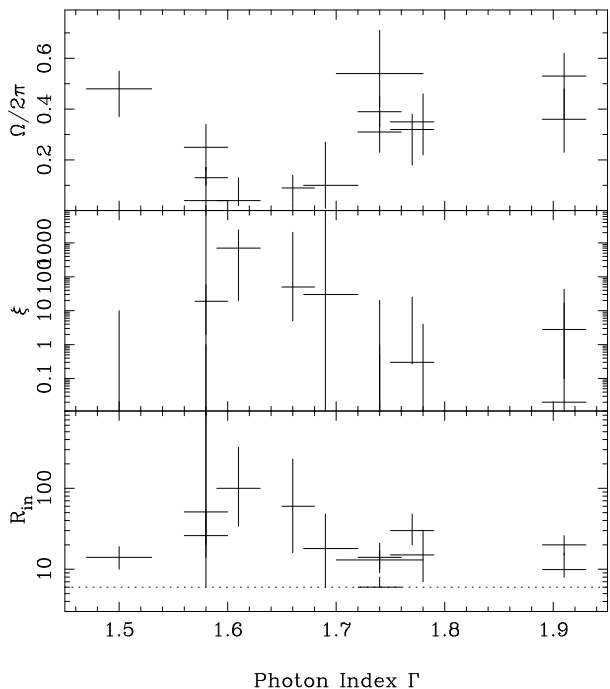}
\caption{
The relativistic reflection spectrum parameters as a function of the intrinsic
spectral index for the model with reflector inclination of $53^\circ$ (table 2). 
The top panel shows the amount of reflection, while the second
panel shows its ionization state and the third shows its inner radius. 
}
\end{figure}

However, this geometrical interpretation is clearly not unique, and many others
have been proposed (see e.g. the review by Poutanen 1998).  If the hot X--ray
emitting plasma forms a (plane parallel) corona above the disk (e.g. Haardt \&
Maraschi 1993) then the reflected spectrum has to cross the corona before
escaping to the observer. A fraction $\exp(-\tau/\cos i)$ (where $\tau$ is the
optical depth and $i$ the inclination of the disk--corona to the observer) will
interact again with the hot electrons in the plasma.  The reflected spectrum
then forms another source of seed photons for Compton scattering, and the
scattered fraction joins the Compton continuum and so is unrecognisable as
reflected flux. Such models were shown to work for Cyg X--1 data up to 20 keV
(Haardt et al 1994), but they predict that there should be a strong break in the
spectrum beyond this energy. Seed photons from the disk are not isotropic, so
the first few Compton scattering retain the imprint of this anisotropy and give
a somewhat flatter spectrum which then joins onto the more isotropic higher
order scatterings (Haardt \& Maraschi 1993). This anisotropy break is {\it not}
present in broad band spectra from Cyg X--1, where the data extend past 20 keV,
which rules out such models (Gierli\'nski et al 1997a).

The lack of an anisotropy break seems to be an insurmountable problem for any
model which has the X--ray emission region above a disk, including the 'active
regions' models which may be more appropriate if the X--ray regions are
energised by localised magnetic reconnection events (e.g. Haardt, Maraschi \&
Ghisellini 1994). However, these models are very attractive given that accretion
disk viscosity is now thought to be driven by magnetic reconnection. But even if
the anisotropy break can be masked (perhaps by the electrons also being
anisotropic or by a distribution of electron temperatures) there are further
problems for models where the X--ray region is above the disk.  If edge effects
are not important (approximately plane parallel geometry) then about half of the
hard X--rays are intercepted by the disk. The majority (whatever is not
reflected) of this flux is thermalised in the disk and re--emerges as soft
photon seeds for the Compton scattering.  These soft photons all have to pass
through the corona in a plane parallel geometry, resulting in strong
cooling. Without a strong anisotropy break then the predicted spectra are too
soft to match the observed low/hard spectra from Cyg X--1 (Dove et al 1997,
Poutanen et al., 1997).  This forces us to the 'active regions' models,
where not all flux of soft photons from the disk will pass through the corona as
edge effects are important. But in these models there is then no overlying
corona that can reduce the amount of reflection seen. If edge effects are
important, then much of the reflection is produced from regions of the disk that
are not underneath the corona.  Reflection is not then supresssed by Compton
scattering in the corona, and there is the strong prediction that we should see
$\Omega/2\pi_{\rm rel}=1$, which is plainly not the case. 

These are serious problems to be faced by all models which have the X--ray
region above the disk. The continuous disk---corona models can reduce the amount
of reflection seen, but predict a large anisotropy break which is strongly ruled
out by the data. Even if there is some way to mask the anisotropy break then
these models produce spectra which are too soft to match what is seen. The
'active regions' models can produce hard spectra, but cannot supress reflection
and again predict an anisotropy break which is not seen. Compared to these
drawbacks, the central source geometry seems an attractive option. 

\section{EQUILIBRIUM ACCRETION FLOW SOLUTIONS}

There are several solutions of the equations for accretion of material onto a
compact object. The most well known of these is that of Shakura \& Sunyaev
(1973; hereafter SS), which is the geometrically thin, optically thick accretion
disk. In this solution, the gravitational energy released is radiated away
locally as (quasi)blackbody radiation. However, other solutions are also
possible: the energy can instead be radiated by Compton scattering in an
optically thin(ish), geometrically thick(ish) flow (Shapiro, Lightman \& Eardley
1976). Alternatively, the energy released need not be efficiently radiated. It
can be carried along with the flow (advected) and swept into the black
hole. These inefficient, advectively cooled flows can either be optically thin
(ADAFs: Narayan \& Yi 1995; Abramowicz et al 1995), or optically thick
(Abramowicz et al 1988), depending on the mass accretion rate. In general the
solutions coexist: at any radius there are several different equilibrium flow
configurations for a given mass accretion rate and viscosity (see e.g. Chen et
al., 1995).

\subsection{Shakura--Sunyaev Accretion}

Why should the inner disk radius change, and not extend down to $6\Rg$ ?  The
`standard' accretion disk model developed by SS gives the flow structure in the
limit when the gravitational energy is released in optically thick material.
The cooling is very efficient (blackbody radiation), so all the energy released
locally can be radiated locally, giving a cool, geometrically thin disk
structure.  This becomes unstable when the total pressure is dominated by
radiation pressure (Shakura \& Sunyaev 1976), giving a natural inner disk
cutoff. For steady state models these predict that 
the inner radius should move outwards as the luminosity increases (as the
radiation pressure dominated region increases: SS),
in conflict with the observed trend in the transient GBHC
to show higher reflected fractions and smaller inner disk radii at higher
luminosities (\zycki\ et al 1997; 1998a,b). 

Another, more general problem with the SS accretion disk structures is that such
models give temperatures of order 1~keV for GBHC at high accretion rates but are
quite unable to explain the observed (hard or soft) power law tail out to
energies $\ge 100$ keV. Either there are parts of the disk flow in a different
configuration to that of SS, or there is some non--disk structure such as a
corona powered by magnetic reconnection (e.g. Haardt, Maraschi \& Ghisellini
1994). However, this latter possibility (a cool disk underneath a plane parallel
or hemispherical X--ray emitting corona) seems to be ruled out by detailed
spectral fitting (see Section 6). Instead it seems that the spectrum implies a
geometrical constraint that the X--ray emitting plasma and disk occupy more or
less separate regions, such as an outer disk and inner (quasi--spherical ?)
X--ray source (Gierli\'nski et al 1997a, Dove et al 1997, Poutanen et al.,
1997).

\subsection{Advective Accretion}

Recently there has been much excitement about the possibility that another
stable solution of the accretion flow may explain the hard X--ray data.  Below a
critical accretion rate, $\dm\le \mdcrit$, a stable, hot, optically thin,
geometrically thick solution can be found if radial energy transport (advection)
is included (see e.g.\ the review by Narayan 1997). The key assumption of these
models is that the protons gain the energy from viscous processes, while the
electrons only acquire energy through interactions (electron-ion coupling) with
the protons. This coupling timescale can be rather slow compared to the
accretion timescale, so protons can be accreted into the black hole, taking some
of the energy with them.  The energy that the electrons do manage to obtain is
radiated away via cyclotron/synchrotron emission (on an internally generated
magnetic field), bremsstrahlung, or Compton scattering of the resultant spectra
of these two processes. In contrast to the SS accretion flow models, there is
no cold disk in the inner regions, so no strong source of soft seed photons for
Compton scattering, hence the resulting X--ray spectra are typically hard.

Such flows were proposed to explain the hard and very faint X--ray spectra seen
from the transient X--ray GBHC in quiescence (e.g. Narayan, McClintock \& Yi
1997), and then extended by Esin, McClintock \& Narayan (1997) to cover the
whole range of luminosity seen in the GBHC transients. As the mass accretion
rate increases from $\dm << \mdcrit$ to $\dm\sim \mdcrit$ the flow density
increases, so the electron--ion coupling becomes more efficient and the fraction
of energy the electrons can drain from the protons increases. This increase in
radiative efficiency continues to $\dm=\mdcrit$, where only $\sim 35$ \% of the
heat energy is advected. At higher $\dm > \mdcrit$ the advective flow collapses
into an SS disk. Esin et al (1997) identify this change from a hot advective
flow to a cool SS disk as the origin of the hard / soft spectral transition seen
in GBHC, and some observational support for this is indeed seen (\zycki\ et al.,
1998a, Cui et al., 1998, Gierli\'nski et al., 1997b).

These models have been specifically applied to Cyg X--1 by Esin et al (1998).
Their calculated spectra approximately match the overall 2--500 keV GINGA--OSSE
data from Cyg X--1 for $\Rin\sim 200\Rg$, but uncertainties 
in their model may extend this to $\Rin\ge 40\Rg$. This is still larger than most
of the constraints from relativistic smearing (table 2). Our data indicate that
the optically thick material extends much further into the gravitational potential
than anticipated by the advective flows. Similar results are found for the
comparison of the models of Esin et al (1997) against the observed hard state
spectra from the GBHC transient source Nova Muscae (\zycki\ et al., 1998a).

\subsection{Shapiro--Lightman--Eardley flows}

An alternative X--ray hot solution to the cool SS accretion flow was discovered
by Shapiro et al., (1976, hereafter SLE). These solutions are
actually very closely related to the ADAFs discussed above in that they assume
that the protons gain most of the energy from viscous processes, and then 
heat the electrons by electron--ion coupling, resulting in an optically
thin(ish), geometrically thick(ish) flow. This solution of the accretion flow
equations merges into the ADAF branch at a critical mass accretion rate
(see Chen et al 1995; Esin et al 1997; Zdziarski 1998). The key difference is
that on the SLE branch, Compton cooling dominates over advection, while the
opposite is true on the ADAF branch. There is no known
possible hot solution for mass accretion rates above the critical one at which
the ADAF and SLE solutions merge. 

In its original form, the SLE solutions were shown to be viscously stable but thermally
unstable (Pringle 1976). An increase in the heating rate is not balanced by an
increase in the cooling because the electron--ion collisons become less
efficient as the temperature increases. The key to the growth of interest in
ADAFs is that they are stabilsed against this due to the dominance of advective
cooling. However, the SLE is not {\it necessarily} thermally unstable. The stability
analyses which have been done generally assumed that the viscous timescale is
long compared to the thermal timescale, so that the density of the region cannot
change during the perturbation.  This is not true for a geometrically thick(ish)
flow. Many stability analyses also neglected advective cooling, and/or external
Compton cooling on seed photons which vary. We urge further study of the thermal
stability of these flows.

Understanding the stability (and hence the existence) of these flows is
important since the SLE composite geometry is rather similar to that inferred
from the data, with a hot, quasi--spherical plasma cooled by radiation from an
external SS accretion disk.  The advantage the SLE solution has over the ADAF is
that it radiates efficiently, which is generally required from the rapid
observed variability of accreting black holes.  However, the disk transition
again occurs where radiation pressure dominates over gas pressure, so these
models again apparently predict that the inner disk radius is larger at high
mass accretion rates, opposite to the observed trend.

\section{UNSTABLE ACCRETION IN AN SS DISK}

The thermal instability above is produced wherever an increase in the mass
accretion rate is not balanced by a corresponding change in the cooling rate. 
Similarly, a disk is viscously unstable if a small increase in the external mass
accretion rate cannot be balanced by the viscosity to produce a corresponding
change in the mass accretion rate through the disk.
These two are related in that flows which are 
viscously unstable are also necessarily thermally unstable.

The ionization of Hydrogen produces a disk transition which is viscously (and
thermally) unstable. Below temperatures of $\sim 5000$ K the opacity is
dominated by molecular lines, while above this temperature there is a marked
increase in opacity due to bound--free ionization of Hydrogen.  However, for
temperatures beyond $\sim 10^4$ K the bound--free opacity drops since hydrogen
is then mostly ionized. This gives a local maximum in opacity at $\sim
6000-10^4$ K (e.g. Hure et al 1994). In this range an increase in $\dm$ gives
rise to a large increase in opacity, trapping the radiation and raising the
central (equatorial plane) temperature of the disk. This in turn raises the
sound speed, $c_s$, and the kinematic viscosity, $\nu=\alpha c_s H$, leading to
an increase in the rate at which the disk material is accreted onto the compact
object (see e.g. the review by Osaki 1996).

If this instability can propagate throughout the disk then this leads to two
distinct states. In quiescence the disk is cool and has low kinematic viscosity
-- generally so low that the rate at which material is added to the disk from
the external accretion source is faster than the rate at which mass can be
viscously transported inwards from the outer disk. Mass builds up at the outer
edge of the disk, and the structure is {\it not} that of a cool steady SS
disk. Eventually, the outer disk mass is so large that hydrogen can be ionized,
and then the disk switches to a hot state where the kinematic viscosity is high and
steady state can be achieved.

This instability is thought to be the trigger for the outburst in the transient
black hole systems (see e.g. Lasota, Narayan \& Yi 1996), though it was first
discovered in the context of the quiescent/outburst behaviour of disk accreting
white dwarfs (see e.g. the review by Osaki 1996).  However, one major problem
with these models is that the accretion rate inferred in quiescence (as evinced
by the X--ray emission) is rather larger than the maximum accretion rate for
which the cool quiescent disk can be sustained if the disk extends all the way
down to the compact object (Lasota et al., 1996). This led to the idea of
a `hole' in the inner disk in quiescence, perhaps formed by evaporation (Meyer
\& Meyer--Hofmeister 1994).  However, another possibility which has yet to be
explored in this context is that the inner disk in quiescence is simply
optically thin. The surface density profile for the non--stationary cool
quiescent disk is peaked to the outside of the disk, so it is very easy for the
inner regions to be optically thin (e.g. Cannizzo 1998, figure 7), especially as
the {\it absorption} opacity of the cool disk is low so even if the photon is
scattered many times, there is still a high probability that it can ultimately
escape (effectively optically thin). This means that the inner disk is unlikely
to be able to thermalise the energy released locally, and so may not radiate
(i.e. cool) efficiently. The material would then heat up, and turn into
something like an advective or externally cooled SLE flow which would appear to
emanate from a (non stationary) optically thick disk.

However, the main problem with such models is that while this may describe the
situation in quiescence, it seems highly unlikely that it has any relevance to
the high mass accretion rates required for Cyg X--1 and the other X--ray bright
transient GBHC. Another instability that can give the same effect but at high
mass accretion rates is required. One obvious possibility is the instability
which occurs when radiation pressure starts to dominate in the disk. The
internal pressure (and hence the sound speed and kinematic viscosity) then rises
very rapidly ($\propto T^4$), and the mass accretion rate through the radiation
pressure dominated part of the disk can then increase faster than the increase
in external mass accretion which triggered the rise in temperature, so leading
to depletion of the inner disk. At higher mass accretion rates the flow is again
stabilised though the excess radiation being carried along with the flow and
advected into the black hole (Abramowicz et al 1988).  Models of such flows are
beginning to be developed (Szuszkiewicz \& Miller 1998) in response to observations
of rapid inner disk instabilities in the superluminal accreting black hole
candidate GRS1915+105 (Belloni et al 1997), but are currently in their
infancy. It remains to be seen whether models of a non--stationary flow inside the
radiation pressure dominated regime could reproduce what is seen in Cyg X--1 and
other low state GBHC spectra.

\section{INHOMOGENEOUS ACCRETION}

It seems highly unlikely that accretion flow can change smoothly between an
optically thick cool solution and an optically thin hot solution. Some
transition region where cool clumps are embedded in hot plasma seems physically
more probable, and turbulence is indeed seen in the numerical calculations of
Abramowicz, Igumenshchev \& Lasota (1998).  Recent work on an inhomogeneous
solution to the accretion flow equations has shown that there is a stable
solution to the radiation pressure dominated SS disk if the majority of mass is
in dense clumps which are optically thick, while the remainder forms an
optically thin(ish), X--ray hot plasma (Krolik 1998), with a physical mechanism
for clumping provided by the photon bubble instability (Gammie 1998). However, one
problem with these models is that they seem to predict that the covering
fraction of cool clumps is close to unity for stellar mass black holes, in
conflict with the observations.

\section{CONCLUSIONS}

We detect significant relativistic smearing of the reflected spectral features
in Cyg X--1, giving the first observational confirmation of an inner accretion
disk in Cyg X--1. Our model calculates the total (line plus continuum) reflected
spectrum for a given ionization state and iron abundance and then applies the
relativistic smearing at a given inclination. These four parameters can be
uniquely determined from X--ray spectra (assuming a given illuminating flux) with
moderate energy resolution, and excellent signal--to--noise data extending
beyond 9 keV.  These conditions are satisfied in Cyg X--1 for the EXOSAT GSPC
and ASCA GIS spectra. The GINGA data are taken in a reduced spectral resolution
mode, while the ASCA SIS data have insufficient signal--to--noise above 9 keV.

For the ASCA GIS and EXOSAT GSPC data the reflected spectrum is constrained to
be from material inclined at $\ge 45^\circ$ which is not highly ionized. The low
ionization state rules out Auger ionization as a mechanism for suppressing the
line with respect to the reflected continuum as proposed by Ross et al (1996).
The iron line is not weak with respect to the reflected continuum -- we infer
twice solar iron abundance -- but the visibility of the line is dramatically
reduced by relativistic smearing.

The observed amount of relativistic smearing is not consistent with a disk
extending all the way down to the last stable orbit around the black hole, but instead
indicates that the disk truncates at some typical (but probably variable) radius of $\sim
10-30\Rg$, with a trend that larger transition radii are associated with harder
X--ray spectra. This mirrors the results from the hard spectra of the transient
GBHC such as V404 Cyg and Nova Muscae (\zycki\ et al 1997; 1998ab), suggesting
that low state can be associated with a lack of optically thick disk at the 
innermost radii. 

The amplitude of the reflected component is significantly less than that
expected from an isotropic source above a semi--infinite disk. Again this result
can be understood in the context of an inner `hole' in the optically thick
material which is filled by the X--ray emitting plasma.  Similar conclusions
about the geometry are also found from looking at the detailed continuum
spectral shape of Cyg X--1, where the hardness of the X--ray spectrum implies
that the hot X--ray plasma is spatially separate from most of the accretion disk,
otherwise it intercepts too many soft photons from the disk and produces an
X--ray spectrum which is significantly steeper than observed (Gierli\'nski et al
1997a; Dove et al 1997; Poutanen et al 1997).

The data then strongly indicate that low state GBHC accretion geometry is that
of an inner X--ray hot, optically thin(ish), geometrically thick plasma, which
cools by Compton upscattering of soft photons from an external cool, optically
thick, geometrically thin disk. Such a composite flow seems most likely to be 
a standard SS cool outer disk, with an inner ADAF or SLE flow. The SLE
flow is probably thermally unstable, so for steady state solutions the ADAF seems to be
preferred. However, we stress that the radiation pressure instability may put the
disk in a regime where the flow is not in steady state, or is inhomogeneous. 

\section{ACKNOWLEDGEMENTS}

We thank Ken Ebisawa and Yoshihiro Ueda for kindly allowing us to use their 
Cyg X--1 ASCA data, Andrew King, Uli Kolb and Juri Poutanen for stimulating
discussions, and Andrzej Zdziarski the referee for useful comments. 
CD acknowledges support from a PPARC Advanced Fellowship.

\end{document}